\documentstyle[prl,aps]{revtex}
\begin{document}
\draft
\twocolumn[\hsize\textwidth\columnwidth\hsize\csname @twocolumnfalse\endcsname

\title{Profile scaling in decay of nanostructures}
\author{Navot Israeli\cite{IsraeliEmail} and Daniel Kandel\cite{KandelEmail}}
\address{Department of Physics of Complex Systems,\\
Weizmann Institute of Science, Rehovot 76100, Israel}
\maketitle

\begin{abstract}
The flattening of a crystal cone below its roughening transition is studied by
means of a step flow model. Numerical and analytical analyses show that
the height profile, $h(r,t)$, obeys the scaling scenario $\partial
h/\partial r = {\cal F}(r t^{-1/4})$. The scaling function is 
flat at radii $r<R(t)\sim t^{1/4}$.
We find a one parameter family of solutions for the
scaling function, and propose a selection criterion for the unique
solution the system reaches.
\end{abstract}

]
\input epsf

In recent years it has become technologically possible to design and
manufacture crystalline nanostructures, which are of tremendous importance
for the fabrication of electronic devices. In many cases, these nanostructures
are thermodynamically unstable, and tend to decay with time.
This phenomenon has triggered experimental and theoretical efforts to try
and understand the decay process\cite{Tanaka,RettoriVillain,Bonzel,Villain,Uwaha,OzdemirZangwill,LanconVillain,UmbachKeefeeBlakely,DubsonJeffers,HagerSpohn,RamanaCooper,FuJohnsonWeeksWilliams,Mullins}. 
Under fairly robust
conditions, the decay of a nanostructure at low temperatures
(below the roughening temperature, $T_R$)
is dominated by the motion of atomic steps on
the surface. Hence, attempts have been made to understand and predict the 
relaxation dynamics of
simple step configurations. 

In this
work we analyze, numerically as well as analytically,
the time evolution of a crystalline cone
formed out of circular concentric steps. 
The decay of other types of nanostructures such as bi-periodic surface
modulations 
has been studied experimentally on Si(001) by Tanaka et al.\ \cite{Tanaka}.
Rettori and Villain\cite{RettoriVillain} studied this problem
theoretically in the case of small amplitude modulations. Our study,
on the other hand, is relevant to large amplitude modulations and in
this sense is complimentary to their work. We find that the height of
the cone decays with time as $h(0)-h(t)\sim t^{1/4}$ and the radius
of the plateau at the top of the cone grows with time as $R(t)\sim
t^{1/4}$.

Consider the surface of an infinite crystalline cone, made out of
circular concentric steps of radii $r_i(t)$, separated by flat terraces.
The step index $i$ grows in the direction away from the center of the cone.
We assume no deposition of any new material, no evaporation and
no transport of atoms through the bulk. To calculate the time dependence
of the radii, we have to solve the diffusion equation for adatoms on
the terraces with boundary conditions at the step edges, taking into account
the repulsive interactions (of the form $G(r_{i+1}-r_i)^{-2}$) between steps.
Using a standard approach to do this \cite{RettoriVillain,BalesZangwill},
we arrived at a set of equations of motion for the step radii. It is convenient
to present these equations in terms of dimensionless radii, $\rho_i$, and
dimensionless time $\tau$: 
\begin{eqnarray*}
\rho_i&=&\frac{T}{\Omega \Gamma} \cdot r_i\;,  \\  
\tau&=&D_sC^0_{eq}\Omega\cdot\left(\frac{T}{\Omega\Gamma}\right)^2\cdot
\left(1+\frac{D_sT}{k\Omega\Gamma}\right)\cdot t\;.
\end{eqnarray*}
$C^0_{eq}$ is the equilibrium concentration of a straight isolated step, $T$
is the temperature, $\Gamma$ is the step line tension, $\Omega$ is the
atomic area of the solid and $D_s$ is the diffusion constant of adatoms
on the terraces. $k$ is a kinetic coefficient associated with attachment
and detachment of adatoms to and from steps. 

The equations of motion in terms of these variables take the form
\begin{eqnarray}
\dot{\rho}_i&\equiv&\frac{d\rho_i}{d\tau}=\frac{a_i-a_{i-1}}{\rho_i}\;, \;\;
\mbox{with}  \label{eq:velocity} \\ 
a_i&=&\frac{\frac{1}{\rho_i}-\frac{1}{\rho_{i+1}}+2g \left(
    \xi_i-\frac{\rho_{i-1}}{\rho_i+\rho_{i-1}}\xi_{i-1}-\frac{\rho_{i+2}}{\rho_{i+2}+\rho_{i+1}}\xi_{i+1}\right)}
{\left(1-q\right)\ln \frac{\rho_i}{\rho_{i+1}} - 
q\left(\frac{1}{\rho_i} + \frac{1}{\rho_{i+1}}\right)} \nonumber \\
\xi_i&=&\left(\rho_{i+1}-\rho_i\right)^{-3}\;,\nonumber 
\end{eqnarray}
where the velocities of the first and second steps are modified to include
interactions only with existing steps. Eq.\ (\ref{eq:velocity}) depends on two
parameters: $g$ and $q$. $g=\frac{T^2G}{\Omega^2\Gamma^3}$
measures the strength of step-step interactions $G$ relative to
the line tension $\Gamma$,  
while the parameter $q=(1+\frac{k\Omega\Gamma}{D_sT})^{-1}$ specifies the
rate limiting process in the system. When $D_s \gg k$ (or $q\rightarrow 1$),
diffusion across terraces is fast and 
the rate limiting process is attachment and
detachment of adatoms to and from steps. When $D_s \ll k$ (or $q\rightarrow 0$),
the steps act as perfect sinks and the rate limiting process is diffusion across
terraces.

\begin{figure}[h]
\epsfxsize=90mm
\centerline{\hspace{7mm} \epsffile{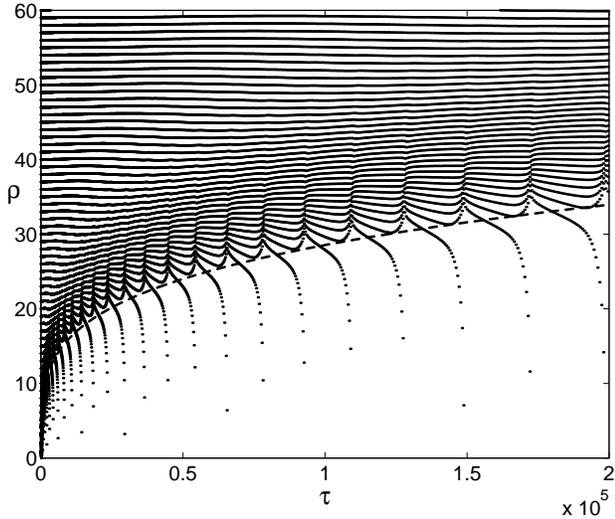}}
\vspace{3mm}
\caption{Time evolution of the step radii in the ADL
  case with $g=0.01$. The front can be fitted by a $\tau^{1/4}$ power law
  (dashed line).}
\label{fig1}
\end{figure}

We integrated Eqs.\ (\ref{eq:velocity}) numerically 
both in the diffusion limited (DL)
and in the attachment/detachment limited (ADL)
cases. When the repulsive interactions
between steps are weak (i.e. $g$ is small), there is a striking difference
between the dynamics in the two limits. In the ADL
case the system becomes unstable towards step bunching, whereas in the DL
case there is no such instability. However, when $g$ is large
enough the instability disappears even in the ADL
case. We limit our discussion to situations where the step bunching instability
does not occur. Fig.\ \ref{fig1} 
shows the time evolution of initially uniformly spaced
steps with unit step separation in the ADL case.
Each line in the figure describes the radius of one step as a function
of time. We note that
the innermost step shrinks while the other steps expand by absorbing the atoms
emitted by the first step. When the innermost step disappears, the next
step starts shrinking and so on. The disappearance time of the $n^{th}$
step, $\tau_n$, grows with $n$ as $\tau_n\sim n^4$.
This process results in a propagating front which leaves a
growing plateau behind it. At large times, the (dimensionless) position of this 
front behaves as
$\rho_{{\mbox \rm front}}(\tau)\sim \tau^{1/4}$. A similar picture arises in the
DL case with differences in the details of the individual
step trajectories. 
This power law is an indication
of a much more
general and interesting phenomenon. It turns out that for large times, 
not only the front position but also the positions of
minimal and maximal step densities scale as $\tau^{1/4}$. In
fact, the step density $D(\rho,\tau)$ obeys the following scaling scenario:
There exist scaling exponents $\alpha$, $\beta$ and $\gamma$ which
define the  scaled 
variables $x\equiv \rho\tau^{-\beta/\gamma}$
and $\theta\equiv\tau^{1/\gamma}$. In terms of these variables  
$D(\rho,\tau)=\theta^\alpha F(x,\theta)$, where the scaling function
$F$ is a {\em periodic} function of $\theta$
with some period $\theta_0$.
Our ansatz is somewhat weaker than standard
scaling hypotheses, which would assume $F$
is independent of $\theta$.
The necessity to introduce a periodic dependence of $F$ on $\theta$
is a manifestation of the discrete nature of the steps. Thus the
disappearance time of step $n$ is $\tau_n=(n\theta_0)^\gamma$.
An immediate consequence of the scaling ansatz is that
if we define $\theta=\bar\theta+n\theta_0$ with $0\leq\bar\theta<\theta_0$,
and plot
$\theta^{-\alpha}
D\left(x\theta^\beta,\theta^\gamma\right)$
against $x$, all the data with different values
of $n$ and the {\em same} value of $\bar\theta$ collapse onto a single curve.

\begin{figure}[h]
\epsfxsize=90mm
\centerline{\hspace{7mm} \epsffile{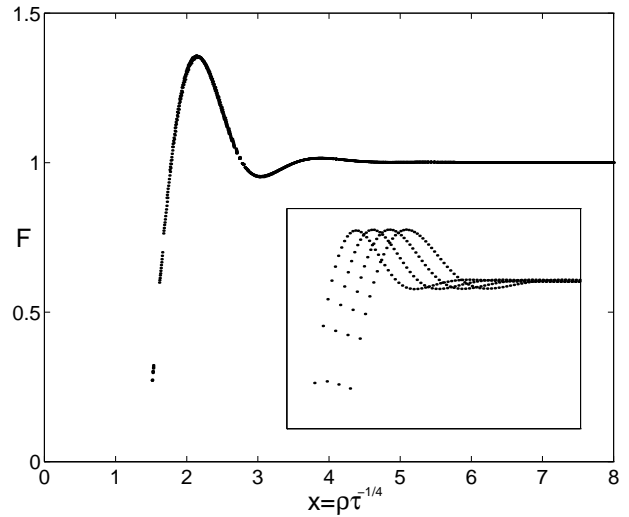}}
\vspace{3mm}
\caption{Data collapse of the density function in the ADL
  case with $g=0.01$. The values of the scaling exponents
  used here are $\alpha=0$, $\beta=1$ and $\gamma=4$. This figure
  shows density functions with 12 
  different values of $n$ and the same value of $\bar\theta$, as a function
  of $x=\rho\tau^{-1/4}$.
  Some of the unscaled data are shown in the inset.}
\label{fig2}
\end{figure}

To verify that our system obeys this scaling ansatz, we define the step
density at a discrete set of points in the middle of the terraces:
\begin{equation}
D\left(\frac{\rho_i(\tau)+\rho_{i+1}(\tau)}{2},\tau\right)=\frac{1}{\rho_{i+1}
(\tau)-\rho_i(\tau)}\;.
\label{eq:discretedens}
\end{equation}
Fig.\ \ref{fig2} is a plot of $D(\rho,\tau)$ as a function of 
$x=\rho\tau^{-1/4}$ for a fixed value of $\bar\theta$
and 12 different values of $n$
in the ADL case.
The excellent data collapse shows that our scaling ansatz indeed holds
with $\alpha=0$, $\beta=1$ and $\gamma=4$.
Data collapse of similar quality is achieved in the DL
case with the same values of the exponents.
The dependence of the scaling function on $\theta$ is very weak in
the DL case, and is more pronounced in the
ADL case.

The above results suggest that the time evolution of the system can be
described by a density, which is a continuous function of both
position and time. In the remainder of this paper we will derive such
a continuum model, carry out a scaling analysis to evaluate the
scaling exponents analytically, and calculate the scaling function. 

Motivated by the
simulation results we assume that the scaling ansatz holds.
This is already sufficient to calculate the values of
$\alpha$ and $\beta$. First, we derive a relation between these 
scaling exponents
by considering the height profile $h(\rho,\tau)$. 
Assuming steps of unit
height, the profile is related to the step density by
\begin{equation}
h(\rho,\tau)=h_0(\tau)-\int_{0}^{\rho}D(\rho',\tau)d\rho'\;,
\label{eq:height}
\end{equation} 
where
$h_0(\tau)$ is the height at the origin. 
Far enough (when $\rho\rightarrow\infty$), $h(\rho,\tau)$ does not change
with time, i.e. $\lim_{\rho \rightarrow \infty}(h(\rho,\tau)-h(\rho,0))=0$.
Combining this with Eq.\ (\ref{eq:height}), 
we arrive at the expression
\begin{equation}
h_0(0)-h_0(\tau)=
\tau^{\frac{\alpha+\beta}{\gamma}}
\int_{0}^{\infty}(F(\infty,0)-F(x,\theta))dx\;, 
\label{eq:hzero}
\end{equation}
where we have used the definition of
the function $F$. 
On the other hand, $h_0(0)-h_0(\tau_n)=n$ because $\tau_n$ is the
disappearance time of the $n^{th}$ step. $\gamma$ satisfies the relation
$\tau_n\sim n^\gamma$, and therefore we have
$h_0(0)-h_0(\tau_n)\sim \tau_n^{1/\gamma}$.
This and the $\tau^{(\alpha+\beta)/\gamma}$ dependence in Eq.\
  (\ref{eq:hzero}) lead to the relation $\alpha+\beta=1$. 

In addition, conservation of the total volume of the system implies
that $\int_0^\infty \rho h(\rho,\tau)d\rho$ is independent of $\tau$.
Integration by parts of the derivative of this integral with respect
to $\tau$ yields the following equation:
\begin{equation}
\int_0^\infty \rho^2\frac{\partial D(\rho,\tau)}{\partial \tau}
d\rho=0\;.
\label{eq:conservation}
\end{equation}
Evaluation of this integral in terms of the function $F$ and
the scaled variables $x$ and $\theta$ shows the integral diverges
unless $\alpha=0$
\cite{long}. Combining this with $\alpha+\beta=1$, we conclude that $\beta=1$.

To evaluate the scaling exponent $\gamma$ and the scaling function $F$,
we continue with
the equation for the full time derivative of the step
density $D$:
\begin{equation}
\frac{dD}{d\tau}=\frac{\partial D}{\partial \tau}+\frac{\partial D}{\partial
  \rho} \cdot\frac{d \rho}{d\tau}\;.   
\label{eq:partials}
\end{equation}
Eq.\ (\ref{eq:partials}) can be evaluated in the middle of the terrace between
two steps (i.e. at $\rho=(\rho_i(\tau)+\rho_{i+1}(\tau))/2$). The l.h.s.
of (\ref{eq:partials}) is calculated by taking the time derivative of
Eq.\ (\ref{eq:discretedens}): $dD/d\tau=-D^2 (\dot{\rho}_{i+1}-\dot{\rho}_i)$.
This together with the fact that
$d \rho/d \tau=((\dot\rho_i(\tau)+\dot\rho_{i+1}(\tau))/2$
leads to the relation 
\begin{equation}
\frac{\partial D}{\partial
        \rho}\frac{\dot{\rho}_{i+1}+\dot{\rho}_i}{2} 
        +\frac{\partial D}{\partial \tau} 
        +D^2(\dot{\rho}_{i+1}-\dot{\rho}_i)=0\;, 
\protect \label{eq:partials2}
\end{equation}
where the step velocities $\dot\rho_i$ can be expressed in terms of the 
$\rho_i$'s using Eq.\ (\ref{eq:velocity}).

Now we change variables to $\theta$ and $x_i\equiv\rho_i\theta^{-1}$,
and transform Eq.\ (\ref{eq:partials2}) into an equation for the scaling 
function $F$. In terms of these variables, Eq.\ (\ref{eq:discretedens}) takes
the form
\begin{equation}
x_{i+1}-x_{i}=\frac{\theta^{-1}}{F\left((x_{i+1}+x_i)/2,\theta
\right)}\;.
\label{eq:xsdifference}
\end{equation}
According to this, the difference between successive $x_i$'s is  
of order $\theta^{-1}$ wherever $F$ does not vanish. 
In the large $\theta$ (long time) limit
these differences become vanishingly small. 
This allows us to go to a continuum limit in the variable $x$,
by expanding all the terms in the equation for the scaling function $F$
in the small parameter $\theta^{-1}$. The final result of these
manipulations is the following differential equation for $F$:
\begin{eqnarray}
-F'\cdot \frac{x}{\gamma}&+&
\theta^{\gamma-4}\left(F'\cdot \frac{A}{2} +
F^2B \right) \nonumber \\
+\frac{\theta}{\gamma}\cdot \frac{\partial F}{\partial
  \theta}&+&O(\theta^{\gamma-5})=0\;.
\label{eq:partials3}
\end{eqnarray}
$A$ and $B$ are known expressions involving $F$,$F'$,$F''$,$F'''$,
$F''''$, where the primes denote partial derivatives with respect to $x$.
The existence of derivatives up to fourth order in this equation 
is a consequence of
the fact that each step ``interacts'' with four other steps (two on each side)
through the equations of motion (\ref{eq:velocity}).
A detailed derivation of the continuum model and the exact expressions
for $A$ and $B$ will be given elsewhere \cite{long}. 

Consider Eq.\ (\ref{eq:partials3}) at large $\theta$. Our expansion in
the small parameter $\theta^{-1}$ is valid only at values of
$x$ where $F$ does not diverge or vanish (see above). 
Therefore, the first term in
Eq.\ (\ref{eq:partials3}) is $O(1)$. This term has to be canceled by
the second term if we require $F$ to satisfy a single differential
equation. Hence, we must have 
\begin{equation}
\gamma=4\;.  \label{eq:exprelation1}
\end{equation}
The fourth term vanishes as $\theta\rightarrow\infty$,
since $\gamma-5<0$ and the
third term must vanish as well. Therefore, in the large $\theta$ limit,
$F$ is only a function of $x$, and we are left with an
ordinary differential equation for $F$: 
\begin{equation}
F'\left(\frac{A}{2}-\frac{x}{4}\right)
+F^2B=0\;. 
\label{eq:differentialeq}
\end{equation}
Let us emphasize that our continuum model is valid for
arbitrarily large surface curvature and slope (unlike other treatments
\cite{RettoriVillain,OzdemirZangwill}). Moreover, since our model is
an expansion 
in the truly small parameter $\theta^{-1}$ (see Eq.\
(\ref{eq:xsdifference})) it becomes {\em exact} in the large $\theta$
(long time) limit. 
Note that in going to the continuum limit we lost the periodic dependence
of $F$ on $\theta$, which is a manifestation of the {\em discrete} nature of the steps.

We now turn to study the solutions of Eq.\ (\ref{eq:differentialeq}). We will
consider only the DL case, but an equivalent treatment can be
applied to more general situations \cite{long}. In the DL case
Eq.\ (\ref{eq:differentialeq}) becomes \cite{HagerSpohn} 
\begin{eqnarray}
\lefteqn{g  \left(
        12F'F'''+3FF''''+9F''^2
        +\frac{15F'F''+5FF'''}{x} \right.} \nonumber \\
        & &  \left.  -\frac{7\left(F'^2+FF''\right)}{x^2}
        +\frac{6FF'}{x^3}-\frac{3F^2}{x^4} \right) 
-\frac{xF'}{4}-\frac{3}{x^4}=0\;.
\label{eq:psdiffeq}
\end{eqnarray}
Without loss of generality, we choose the boundary condition at
infinity to be $F(\infty)=1$. Any other choice is equivalent to our
choice with a different value of the interaction parameter $g$,
since Eq.\ (\ref{eq:psdiffeq}) is invariant under the transformation
$F\rightarrow aF$, $x\rightarrow a^{-1/4}x$ and
$g\rightarrow a^{-2}g$. Numerical solutions of Eq.\
(\ref{eq:psdiffeq}) which satisfy $F(\infty)=1$ indicate that there
exists a point $x=x_0$ near which $F$ behaves as $F(x)\sim\sqrt{x-x_0}$.
An analytical expansion of Eq.\ (\ref{eq:psdiffeq}) for small $F$
also leads to the same conclusion. Thus, our model naturally predicts
a singular point in the profile at which $F\rightarrow 0$. We can
prove \cite{long} that $x_0$ is the scaled position of the boundary of
the plateau at the top of the hill.

In the derivation of Eq.\ (\ref{eq:psdiffeq}) we assumed $F(x)\neq 0$. This is
clearly violated on the plateau. Eq.\
(\ref{eq:psdiffeq}) is therefore valid only at $x>x_0$, while
for $x<x_0$, the solution for $F$ is simply $F=0$. Now we have to 
solve Eq.\ (\ref{eq:psdiffeq}) for $x>x_0$ with some boundary
conditions at $x=x_0$ together with the condition $F(\infty)=1$.
In addition, we have to make sure that all the atoms expelled by the
growing plateau at $x<x_0$ are absorbed by the steps at $x>x_0$.
This is taken care of by the conservation law (\ref{eq:conservation}).
We rewrite (\ref{eq:conservation}) in terms of scaled variables and get
the following equation:
\[\int_0^\infty x^2\left(F(x)-1\right)dx=
\int_{x_0}^\infty x^2\left(F(x)-1\right)dx-\frac{x_0^3}{3}=0\;.\]
To carry out the last integral 
we integrate Eq.\ (\ref{eq:psdiffeq}) multiplied by $x^2$ from
$x_0$ to $\infty$, and obtain the boundary condition
\begin{eqnarray}
\lefteqn{\lim_{x \rightarrow x_0}  
\left\{\vphantom{\frac{3}{x}} g  \left( 6FF'
+xFF''+xF'^2 \right. \right.} \nonumber \\
& & \;\;\;\;\;\;\;\; \left. \left. -3x^2FF'''-9x^2F'F'' 
                                        \right) 
                                -\frac{3}{x} \right\}=0\;.
\label{eq:boundry2}
\end{eqnarray}

\begin{figure}[h]
\epsfxsize=90mm
\centerline{\hspace{7mm} \epsffile{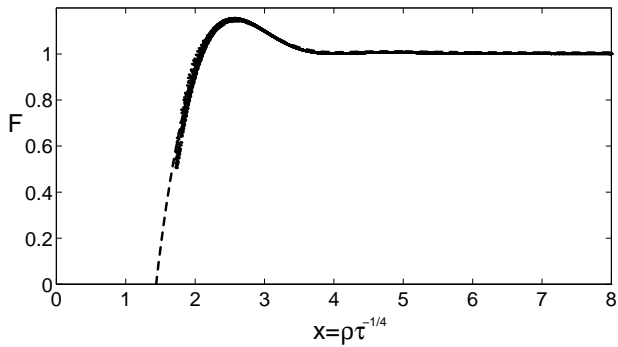}}
\vspace{3mm}
\caption{Data collapse of the density function in the diffusion limited case
  with $g=0.01$. Simulation data from several collapse
  periods are shown as dots, while the dashed line is the
$x_0=x_0^*$ solution of Eq. (\protect \ref{eq:psdiffeq}).
  }
\label{fig3}
\end{figure}
So far, we have not determined the value of $x_0$. In fact, 
we solved Eq. (\ref{eq:psdiffeq}) numerically and found a {\em family} 
of solutions
satisfying the boundary conditions, which differ in the value of $x_0$.
For $x_0<x_0^*\approx 1.44$ the equation does not have a solution.
However, for any value of $x_0\geq x_0^*$ there is a single solution
that satisfies the boundary conditions. Despite the existence of many
solutions, our simulations indicate that 
the system reaches a unique scaling solution independent of initial 
conditions.
Fig.\ \ref{fig3} shows an impressive
agreement between the $x_0=x_0^*$ solution 
and the data collapse of density
functions taken from the simulations in the DL case.
Thus, the system dynamically selects the scaling state with the minimal
value of $x_0$. The precise nature of this selection mechanism is not
yet understood and will be investigated in the future.

In summary, we have presented 
a complete description of the relaxation process of an
infinite crystalline cone below it's roughening transition. The hypothesis
that in the long time limit the step density exhibits scaling leads to
an accurate continuous model for the
morphological evolution of the crystal. Using the model,
we were able to derive the exact scaling exponents and
the differential equation that describes the scaling function. 
This equation admits many solutions, and the system dynamically selects
the one with the smallest plateau. We hope this work will motivate
new experiments in which our predictions will be tested.

This research was supported by grant No. 95-00268 from the
United States-Israel Binational Science Foundation (BSF), Jerusalem, Israel.
D. Kandel is the incumbent of the Ruth Epstein Recu
Career Development Chair.


\begin{thebibliography}{99}

\bibitem[*]{IsraeliEmail}
E-mail: israeli@wicc.weizmann.ac.il

\bibitem[**]{KandelEmail}
E-mail: fekandel@wicc.weizmann.ac.il

\bibitem{Tanaka}
S. Tanaka, N. C. Bartelt, C. C. Umbach, R. M. Tromp and J. M. Blakely,
Phys. Rev. Lett. {\bf 78}, 3342 (1997).

\bibitem{RettoriVillain}
A. Rettori and J. Villain, J. Phys. France {\bf 49}, 257 (1988).

\bibitem{Bonzel}
K. Yamashita, H. P. Bonzel and H. Ibach, Appl. Phys. {\bf 25}, 231 (1981); 
H. P. Bonzel and E. Preuss, Appl. Phys. A {\bf 35}, 1 (1984);
H. P. Bonzel and W. w. Mullins, Surf. Sci. {\bf 350}, 285 (1996). 

\bibitem{Villain}
J. Villain, Europhys. Lett. {\bf 2}, 531 (1986).

\bibitem{Uwaha}
M. Uwaha, J. Phys. Soc. Jpn. {\bf 57}, 1681 (1988).


\bibitem{OzdemirZangwill}
M. Ozdemir and A. Zangwill, Phys. Rev. B {\bf 42}, 5013 (1990).

\bibitem{LanconVillain}
F. Lancon and J. Villain, Phys. Rev. Lett. {\bf 64}, 293 (1990).

\bibitem{UmbachKeefeeBlakely}
C. C. Umbach, M. E. Keeffe and J. M. Blakely, J. Vac. Sci. Technol. A
{\bf 9}, 1014 (1991).

\bibitem{DubsonJeffers}
M. A. Dubson and G. Jeffers, Phys. Rev. B {\bf 49}, 8347 (1994).

\bibitem{HagerSpohn}
J. Hager and H. Spohn, Surf. Sci. {\bf 324}, 365 (1995). The straight
steps limit of our DL case coincides with the
continuum theory of these authors. 

\bibitem{RamanaCooper}
M. V. Ramana Murty and B. H. Cooper, Phys. Rev. B {\bf 54}, 10377
(1996). These Monte Carlo simulations agree with \cite{RettoriVillain,OzdemirZangwill}.

\bibitem{FuJohnsonWeeksWilliams}
E. S. Fu, M. D. Johnson, D. J. Liu, J. D. Weeks amd E. D. Williams,
Phys. Rev. Lett. {\bf 77}, 1091 (1996).



\bibitem{Mullins}
W. W. Mullins, J. Appl. Phys. {\bf 28}, 333 (1957); {\bf 30}, 77 (1959).

\bibitem{BalesZangwill}
G. S. Bales and A. Zangwill, Phys. Rev. B {\bf 41}, 5500 (1990).

\bibitem{long}
N. Israeli and D. Kandel, unpublished.

\end{thebibliography}
\end{document}